# Inter Time Series Sales Forecasting

Mrs. Manisha Gahirwal
Information Technology, Vivekanand Education Society's Institute of Technology,
Chembur, India
manishavesit@gmail.com

Vijayalakshmi M.
Information Technology, Vivekanand Education Society's Institute of Technology,
Chembur, India
viji.murli@gmail.com

*Abstract—* **Combining forecast from different models has shown to perform better than single forecast in most time series. To improve the quality of forecast we can go for combining forecast. We study the effect of decomposing a series into multiple components and performing forecasts on each component separately... The original series is decomposed into trend, seasonality and an irregular component for each series. The statistical methods such as ARIMA, Holt-Winter have been used to forecast these components. In this paper we focus on how the best models of one series can be applied to similar frequency pattern series for forecasting using association mining. The proposed method forecasted value has been compared with Holt Winter method and shown that the results are better than Holt Winter method**

*Keywords- Association mining, Combining, Decomposition, Forecasting, Inter time series.*

## I. INTRODUCTION

**Forecasting** is the process of making statements about events whose actual outcomes have not yet been observed [2, 7, and 14]. It is a method or a technique for estimating future aspects of a business or the operation. Forecasts are important for short-term and long-term decisions. Businesses may use forecast in several areas like technological forecast, economic forecast, demand forecast etc.

A sales forecast is a prediction based on past sales performance and an analysis of expected market conditions. [2, 4, 5] Many management and control decisions are often influenced by the current market situation and on how it is expected to change in the near future.

Sales forecasts help investors make decisions about investments in new ventures. They are vital to the efficient operation of the firm and can aid managers on such decisions as the size of a plant to build, the amount of inventory to carry. Building a decision support system for forecasting is generally non-trivial due to the complex factors affecting demand. Forecasting methods used in practice vary from domain to domain. Here we restrict ourselves to time series forecasting [10, 11, 12, and 18].
The challenges are:
- Decreasing the error in the forecast as much as possible
- Finding / developing a relatively inexpensive easy to maintain forecasting system that guarantees desired accuracy.

An accurate prior estimate of the possible changes in the key factors can go a long way in improving the quality of decision taken. Sales forecasting is a self-assessment tool for a company. You keep taking the pulse of the company to know how healthy it is. Timely and accurate sales forecast are crucial in bridging the gap between supply and demand. A good forecasting system helps to decrease inventory holding costs while at the same time reducing the probability of stock-outs and customer wait times.

There are two main approaches to forecasting- Quantitative methods (objective approach) and Qualitative methods (subjective approach) [14].

***Quantitative forecasting methods*** are based on analysis of historical data and assume that past patterns in data can be used to forecast future data points. There are two types of quantitative methods: ***Times-series method*** and ***explanatory methods***. Quantitative methods include statistical, state space models and those based on machine learning techniques such as neural networks genetic algorithms etc. Statistical techniques use classical regression or time-series models such as exponential smoothing and ARIMA (Auto-Regressive Integrated Moving Average). The experiments reported in this paper exclusively use times series models [2, 14].

***Qualitative forecasting techniques*** employ the judgment of experts in specified field to generate forecasts. They are based on educated guesses or opinions of experts in that area.

A ***time series*** **[5, 6, 10, 11, 12, and 14]** is a sequence of data points, measured typically at successive times spaced at uniform time intervals. Time series analysis comprises methods for analyzing time series data in order to extract meaningful statistics and other characteristics of the data [10, 12]. **Time series** *forecasting* is the use of a model to forecast future events based on known past events to predict data points before they are measured. Time series are very frequently plotted via line charts. E.g. Stock market, sales forecast, here time series analysis is applicable.

Time-series methods make forecasts based solely on historical patterns in the data. A first step in using time-series approach is to gather historical data. The historical data is representative of the conditions expected in the future.

Here we are concentrating on sales data. Sales forecasting is an important part of supply chain management both at the retailer end and at the distributors, manufacturers and suppliers.

Forecasting problem can be stated as predicting future values $x_{t+h}+h$, given a part of a time series $x_t, x_{t-1}, \cdots, x_{t-w+1}$ (called pattern), where w is the length of the window on the time series and h _ 1 is called the prediction horizon.

To obtain these predictions, normally statistical models such as ARIMA(Auto-regressive Integrated moving Average) models, exponential smoothing models tec. are used these models are typically trained by using historical data and can then be used to get the forecast for future points. The main aim here is to improve the forecasting accuracy. Many techniques such as decomposition, combining are used for this purpose [2, 4].

One approach to forecasting involves choosing a single model at each point, typically based on past performance. However, model selection is associated with the instability problem whereby even a slight change in the data sometimes results in the choice of a different model. Another approach is the use of a 'Combination Forecast' [4, 14] where, instead of trying to choose the single best model, we attempt to identify a group of models which, in conjunction, help to improve forecast accuracy.

## II. RELATED WORK

*A time series* is a sequence of observation taken sequentially in time. It can be seen as being composed of three components. Trend (Change in the time series), Seasonality (repetition of a particular pattern of observations after certain fixed time interval) and Irregular component (Random noise). These three components are used for decomposition methods that will generate individual components series from original series. These methods form one of the important parts of the literature on forecasting. There exist various decomposition methods that will generate individual components series from the original series. These methods form one important part of the literature on forecasting. There exist various decomposition methods e.g. Holt-Winter method, Exponential Smoothing method, Exponential smoothing with a damped multiplicative trend that are used to get the component series from the original series [14].

Here we covered an issue in the literature on forecasting is the problem of model selection for combining forecast[2,4], normally there are many forecasting experts available so one has to decide whether to select only a single expert or to combine the forecasts of different experts in one way to get the final forecast

In recent times there has been an increasing interest in using data mining - based methods to identify the best combination of models that can increase forecast accuracy. In this paper, we use Association mining [1,3,5,7,8,9,12,13,16]algorithm to find best models to calculate final forecast values.

A. *Forecasting methodology :*

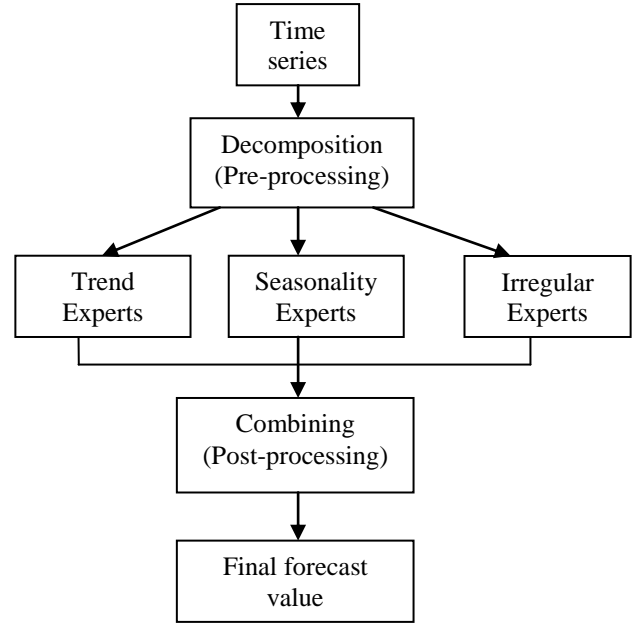

Figure 1. Time series forecasting model

B. *Forecasting concept and Techniques*

The main objective of the time series analysis is to model a process, which is generating the data, to provide compact description and to understand the generating process. Decomposition (pre-processing) and combining (post-processing) when applied on the time series result in improvement of the accuracy of forecasts. After decomposing the series statistical experts such as ARIMA [2, 14, and 15] are applied to get the forecasts of component series.

*Forecasting Models*

Forecasting models are the mathematical models which are makes certain assumptions about the time series, captures certain features and then predicts the next unknown value. Some mathematical models, which is used is given below

*Stochastic Models:*

1) *Holt-Winter Method :*

The Holt-Winter's method is one of the best known forecasting techniques for the time series that has both trend and seasonal components. There are two variants of this method, additive and multiplicative. The seasonality is multiplicative if the magnitude of the seasonal variation increases with an increase in the mean level of the time series. It is additive if the seasonal effect does not depend on the current mean level of the time series. The basic Holt-Winter forecasting method with multiplicative seasonality (exponential smoothing of level (St), trend (Tt) and seasonal index (It) ) is described by

$$S_t = (D_t / I_{t-p}) + (1- )(S_{t-1} + T_{t-1})$$

$$T_t = (S_t - S_{t-1}) + (I - )T_{t-1}$$

$$I_t = (D_t / S_t) + (1 - )I_{t-p}$$

Here p is the number of observation points in a cycle (p = 4 for quarterly data) are the smoothing constants.
The forecast at time t for time t + i is:
$$D_{t+i} = (S_t + i \times T_t )I_{t-p+i}$$

*2) Autoregressive Integrated Moving Average (ARIMA)*

Models for time series data can have many forms. When modeling variations in the level of a process, three broad classes of practical importance are the *autoregressive* (AR) models, the *integrated* (I) models, and the *moving average* (MA) models. These three classes depend linearly on previous data points. Combinations of these ideas produce autoregressive moving average (ARMA) and autoregressive integrated moving average (ARIMA) models.

The Autoregressive Integrated Moving Average (ARIMA) models [2], or Box-Jenkins methodology, are a class of linear models that is capable of representing stationary as well as non-stationary time series. ARIMA models rely heavily on autocorrelation patterns in data.

The Box- Jenkins ARIMA models are most general class of models for forecasting a time series which can be done by transformations such as differencing and lagging.

ARIMA methodology of forecasting is different from most methods because it does not assume any particular pattern in the historical data of the series to be forecast. An ARMA model predicts the value of the target variable as a linear function of Lags of differenced series appearing in forecasting equation lag values (the auto-regressive part) plus an effect from recent random values (the moving average part). A time series which needs to be differenced to be made stationary is said to be an "integrated" version of a stationary series.

Autoregressive (AR) models can be coupled with moving average (MA) models to form a general and useful class of time series models called *Autoregressive Moving Average (ARMA)* models. These can be used when the data are stationary. This class of models can be extended to non-stationary series by allowing the differencing of the data series. These are called *Autoregressive Integrated Moving Average (ARIMA)* models.

Seasonal ARIMA models a pattern that repeats seasonally over time. It is classified as an **ARIMA (p,d,q) x(P,D,Q)** model,

  P=number of seasonal autoregressive (SAR) terms,
  D=number of seasonal differences,
  Q=number of seasonal moving average (SMA) terms

Non Seasonal ARIMA Models are classified as an "ARIMA (p,d,q)" model.

Where, **p** is the number of autoregressive terms,
**d** is the number of non seasonal differences, and
**q** is the number of lagged forecast errors in the prediction equation.

*C. Decomposition of Time Series*

**A time series** can be broken down into its individual components. The decomposition of time series is a statistical method that breaks a time series down into its components (Trend, Seasonal, Cyclical, and Random / Irregular)

Decomposition methods usually try to identify two separate components of the basic underlying pattern that tend to characterize economics and business series. The trend Cycle represents long term changes in the level of series. The Seasonal factor is the periodic fluctuations of constant length that is usually caused by known factors such as rainfall, month of the year, temperature, timing of the Holidays, etc.

Any time series is a composition of many individual underlying component time series. Some of these components are predictable whereas other components may be almost random which can be difficult to predict. Decomposing a series into such components enables us to analyze the behavior of each component and this can help to improving the accuracy of the final forecast. A typical sales time series can be considered to be a combination of four components i.e. trend component ($T$), cyclic component(C), seasonal component(S) and irregular component (I).

*a. Decomposition Models :*

Mathematical representation of the decomposition approach is:

$$D_t = (S_t, T_t, IC_t)$$

Where,
$D_t$ is the time series value (actual data) at period t.
$S_t$ is the seasonal component ( index) at period t.
$T_t$ is the trend cycle component at period t.
$IC_t$ is the irregular (remainder) component at period t.

The **Seasonal** component models patterns of change in a time series within a year. These patterns relate to periodic fluctuations of constant length, tend to repeat themselves each year. **The Trend** represents changes in the level of the series. The **Cyclic** component refers to patterns, or waves, in the data that are repeated after approximately equal intervals with approximately equal intensity, with period normally larger than seasonal period. Usually the trend and cyclic component are together treated as the **Trend** component. The **Irregular** component refers to variations not covered by the above. (Residuals)

The mathematical representation for decomposition is

*Trend (T)*

$$T_t = \begin{cases} (D_1 + D_2 + \dots + D_{12})/12 & \text{for } t = 1 \text{ to } 12 \\ (D_{t-11} + D_{t-10} + \dots + D_{t-1} + D_t)/12 & \text{for } t > 12 \end{cases}$$

*Seasonality (S)*

$DT_t = D_t / T_t$

$S_t = Average\ (DT_t, DT_{t-12}, DT_{t-24}, \dots, DT_{t-k})$ till $t - k > 0$

*Irregular Component (IC)*

$IC_t = DT_t / S_t$

Forecasts of the individual components can be combined to get the final forecast by using operator '+' and 'x' so there are two models shown below,

The decomposition could be additive if the magnitude of seasonal fluctuations do not vary with the level of the series.

*Dt = Tt x St x It  --------- Multiplicative model*

We have a multiplicative decomposition if seasonality fluctuates and increases and decreases with the level of the series.

*Dt = Tt + St + It  ---------- Additive model*

*Multiplicative model* is more prevalent with economic series since most seasonal economic series have seasonal variation which increases with the level of the series. Further experiments carried out on a sample set of sales series has indicated that seasonal multiplicative model performs better than the additive model. In this research we shall thus use the multiplicative decomposition model.

### D. Combine method Forecast :

There are many forecasting models used in time series forecasting. Typically one model is selected based on a selection criterion, hypothesis testing and/ or graphical inspection and selected model is used for forecasting. However, model selection is often unstable and may cause an unnecessary high variability in the final prediction. To overcome the above problems, combining techniques can be useful. In combining, we combine the forecast from different experts to generate a final forecast. This combining function can be as simple as taking mean to a sophisticated as applying dynamic function over dynamic set of forecasting experts.

The ideas of combining forecasts implicitly assume that one model could not identify the underlying process, but different forecasting models could capture different aspects of the information available for prediction. Combining forecasts is used for risk minimization as it reduces the variance of the final forecast. Combining techniques is also used to improve forecast accuracy under certain circumstances.

Selection of experts to be participating in combining is important if we want to improve forecast accuracy. There are two most popular and simple methods to combine forecasts. One is taking mean and another is taking median of forecasts from different forecasting methods. Out of these two methods many times median outperform mean methods. Combining is more useful for long-range forecasting because of the greater uncertainty. An alternative viewpoint is that random errors are more significant for short-range forecasts, because these errors [14, 15] are off-setting and a combined forecast should reduce the errors.

### E. Error Measures :

The forecast error is the difference between the actual value and the forecast value for the corresponding period (t).

Forecast Error (Et) = Actual value (Yt) − Forecasted value (Ft)

Error measure has an important role in calibrating a refining forecasting model / method. As the main aim in forecasting is to increase the accuracy of the forecasts, error measures are very important from a forecaster's point of view. There are varieties of error measures that are used in practice. Some of them are listed below:

1. **Mean Squared Error (MSE) :**

$$MSE = \frac{\sum_{t=1}^{N} E_t^2}{N}$$

2. **Mean Absolute Percentage Error (MAPE) :**

$$MAPE = \frac{\sum_{t=1}^{N} |\frac{E_t}{Y_t}|}{N}$$

3. **Mean Absolute Error (MAE) :**

$$MAE = \frac{\sum_{t=1}^{N} |E_t|}{N}$$

4. **Root Mean squared error (RMSE):**

$$RMSE = \sqrt{\frac{\sum_{t=1}^{N} E_t^2}{N}}$$

5. **Median Absolute Percentage Error (MdAPE) :**
   This measure is almost similar to the MAPE, but in MAPE mean is used for summarization where as in MdAPE median is used for summarization across series.

   For the experiments that will be discussed in the coming sections MAPE is used as Error Measure.

### III. ASSOCIATION MINING

Time series Association rule mining [1], has been effective in providing useful information in various domains. Association rule mining, one of the most important and well researched techniques of data mining, was first introduced in. It aims to extract interesting correlations, frequent patterns, associations or casual structures among sets of items in the transaction databases or other data repositories. Frequent model mining is a form of association mining.

An **itemset** is a subset of the items of interest in the database. The **support** of an itemset, X, denoted supp(X), is the frequency of occurrence of X in the database

There are many association mining algorithms, such as Apriori algorithm, Eclat algorithm, FP-Growth and CHARM algorithm etc. But some of the association mining algorithm are not working or very difficult and time consuming to work with huge data set. So CHARM is performing better and faster than the other algorithm.

*A. Consistent Model Mining (CMM) :*

CMM identifies co-occurrences concepts in a huge dataset. It has traditionally been used to capture customer purchasing behavior by monitoring which items frequently occur together in transactions. In our application, an item is an expert. A transaction corresponds to a point in our sales series of interest. The items in the transaction are a list of experts that form the top performing forecasters at that point based on some pre-defined measure of performance. Our goal is to identify one or more sets of experts which frequently perform well across part of a series. Having learned such a set, we use that set of experts to forecast the rest of the series.

IV. OUR APPROACH

In this chapter we will start with the description of Time Series that we have used and Expert Pool what we have made. Then a brief description of the basic decomposition method used in our work. Description of steps used to construct various expert pools. The next section discuss the Advanced association Mining algorithm applied to a text file to get Best and Bad models using which combined forecasted values has been calculated and MAPE has been achieved. Using the 30 time series, which are described, there forecasted value MAPE is compared with HW MAPE and shown that our results are better than HW method. In the last section we have discussed about inter time series [16, 17] and shown that how the best models of one series are giving the same result for another time series of same frequency.

**Step I. Datasets**
**Time series used:**

For our forecasting experiments, we used 30 series from the time series data library. Most of the series represented sales of fast moving consumer goods. The time series data used in this work are monthly sales data. They posses seasonality with the period of 12 month. Some of them are shown below. The details of the series are given in Appendix.

All our experiments were conducted on the above datasets. Each experiment used the firs 70% data values as history and the forecasting started for remaining 30% data.

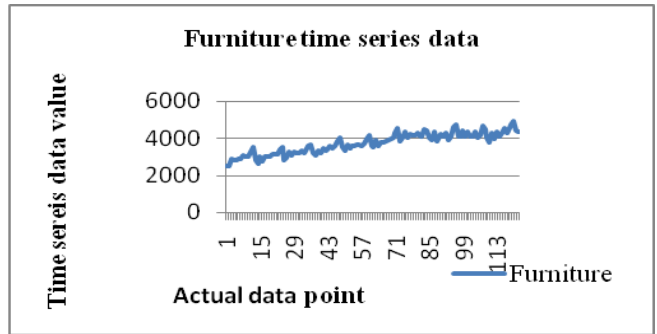

Figure 2. Graph representation of original furniture series

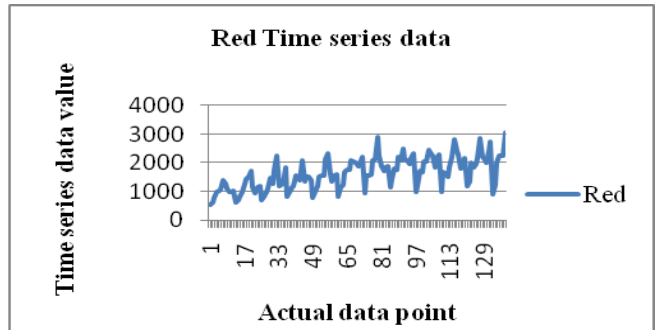

Figure 3. Graph representation of original Red series

**Step II. Decomposition of Time series.**

The time series is actually divided into three components, Trend, Seasonal and Irregular components. So the values for these are calculated by the above explain formulas of decomposition model.

The decomposition could be additive or multiplicative, for out project we have used multiplicative decomposition model.

$D_t = T_t \times S_t \times I_t$ --------- Multiplicative model

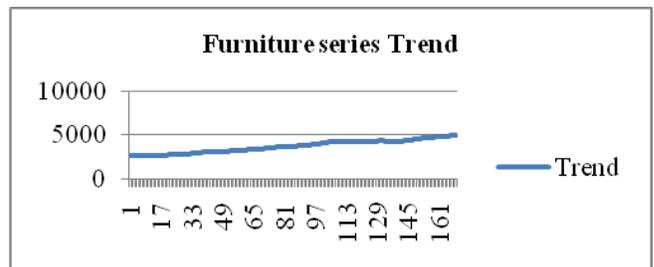

Figure 4. Furniture series Trend

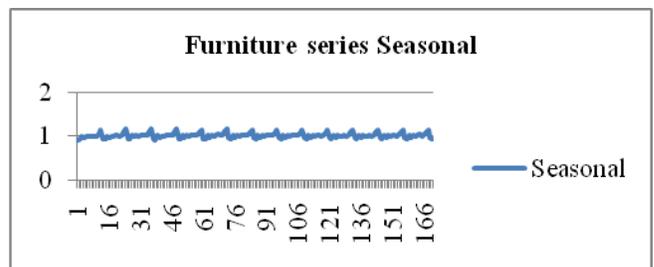

Figure 5. Furniture series Seasonal

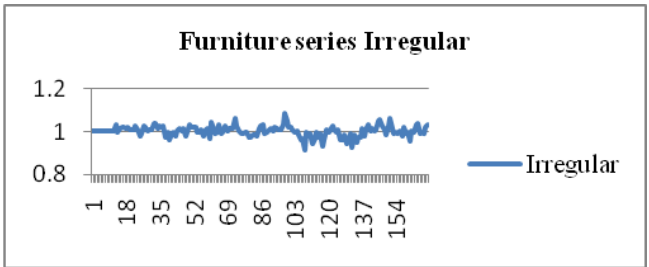

Figure 6 . Furniture series Irregular

**Step III. Using Experts calculate predicated values:**

A method used in forecasting a single component is referred to as an atomic forecaster. A forecaster for the original series is a triplet made up of the atomic forecasters for each component. The set of such triplets is the Cartesian product of the sets of forecasters for the T, S and I components. Each such triplet of atomic forecasters (T, S, I) is called an "expert".

In this work, we use a total of 86 Trend models (atomic forecasters), 33 Seasonal models and 34 Irregular component models. These are mostly ARIMA and seasonal ARIMA models of different orders. The Cartesian product of the Trend, Seasonal and Irregular models gives rise to 96,492 expert forecasts per point. The Appendix includes a list of atomic forecasters used in this work. Two of the best known methods of forecasting seasonal data (such as retail sales) are the Holt-Winter method and seasonal ARIMA .Here we have used the SPSS tool to achieve forecasting values using Trend, Seasonal and Irregular component.

SPSS Forecasting enables analysts to predict trends and develop forecasts quickly and easily — without being an expert statistician.

**Step IV. Filtered Experts Mining:**

The method Filtered Expert Mining generates the set of "good" Trend T, Seasonality S and Irregular I experts for a series. In a similar fashion it was also possible to extract the "bad" T, S and I experts. These bad experts were filtered out of from the set of experts and the surviving experts used for forecasting. This approach was called Fine Grained Filtered Experts approach. Using this approach we found the good and bad experts for a series using just 50% and 70% of the initial points. The forecast accuracy by using the surviving set of T, S, and I experts in combination increased over Holt Winter accuracy.

  a.  **Filtered Expert Mining Algorithm :**

We have achieved forecasted values using ARIMA Model, and then we had calculated 96,492 experts per point. The initial points of the series are used as the training set, 'n' is the training size.

1. At point (i) {1<=i<=n} take top 20000 experts based on APE,
   APE = (forecast value – actual value) / actual value
2. We sorted those experts according to their APE and found out the top 20,000 and bottom 20,000 experts.
3. Calculated the Trend, Seasonal and Irregular experts count and fount the Best and Bad Trend model, Seasonal model and Irregular model by using the following way,
   Best Model where count of (T,S,I) is greater than (20,000/Experts(T,S,I)
   e.g. Trend Best Models = Trend count > (20,000/86)
   Seasonal Best Model = Seasonal count > (20,000/33) and
   Irregular Best Models = Irregular expert count > (20,000/34)
4. Using any standard frequent pattern mining, generate the set of experts that have consistently occurred in the top 20,000 experts list. We get 3 sets of consistent good experts.
5. In the similar way starting with the bottom 20,000 experts at each point it is possible to use 'n' point as training set and identify the set of bad atomic experts for the series, Bad Trend Experts (BadT), Bad Seasonal Experts (BadS), and Bad Irregular Experts (BadI).

**Step V: Association Mining to find Frequent / Consistent Expert**

Here we are mining frequent models using Association Rule mining. We have already fount Best models for each point and as we are working with 70 % and 50% percent on historical data, so we will be using 70% or 50% actual point experts to find best experts with respect to Trend, Seasonal and Irregular.

Apriori algorithm is not efficient for huge dataset so we are using here the CHARM [3] algorithm which is implemented by Zhaki, with modification.

It is difficult to handle a huge dataset, because there are problems for finding long patterns. So

There are two current solutions to the long pattern mining problem. The first one is to mine only the maximal frequent itemsets, which are typically orders of magnitude fewer than all frequent patterns. While mining maximal sets help understand the long patterns in dense domains, they lead to a loss of information; since subset frequency is not available maximal sets are not suitable for generating rules The second is to mine only the frequent closed sets. Closed sets are lossless in the sense that they uniquely determine the set of all frequent itemsets and their exact frequency. At the same time closed sets can themselves be orders of magnitude smaller than all frequent sets, especially on dense databases

**Working of CHARM Algorithm**

For finding patterns we have used the basic CHARM algorithm, then we exteded the algorithm to find the unique experts.

**Advanced CHARM Algorithm:** The above algorithm gives us frequent patterns which are having repetitive models.

So we separated those models and selected only the unique and consistent models for getting Best and T, S, I models. Result achieved after using the advanced CHARM Algorithm is applied to get forecasted values without using all the experts of Trend, Seasonal and Irregular but the consistent experts has been used and applied on the ~30% data of actual time series [10,11] to verify the forecasted values .

**Step 5: Techniques of Combining**

There are two dimensions of the combing problem that are identified. The following techniques of combining are considered here:
- Use mean/ median of all experts.
- Best Method is to use best expert at each point of time. One criteria of goodness at a particular time is the MAPE measure at that time.

Combining experts can be performed by selecting the subset of experts at each point of time and take their mean as a final forecast. Here we eliminate altogether most of the experts and combine uniformly the remaining experts.

The results of two series for 24 points are shown below to show the result of combined forecast values and actual value of that point.

Graph for Abraham Series and Merchandise time series is shown below

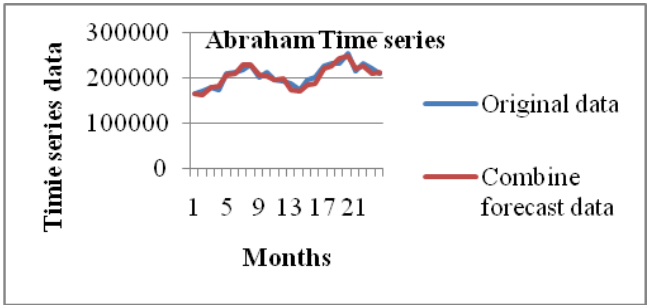

Figure 7. Original and Forecast value for Abraham time series.

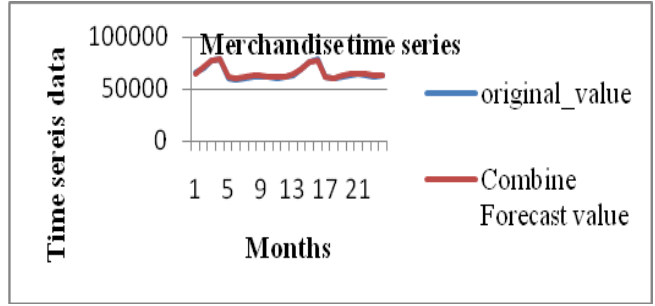

Figure 8. Original and Forecast value for merchandise time series.

**Step 6: Comparison of calculated consistent Model Mining MAPE with Holt Winter MAPE**

To evaluate the accuracy of our algorithm we recorded the MAPEs obtained for the last 24 points of each series, and compare it with the MAPE of the Holt Winter forecast of the last 24 points of the series.

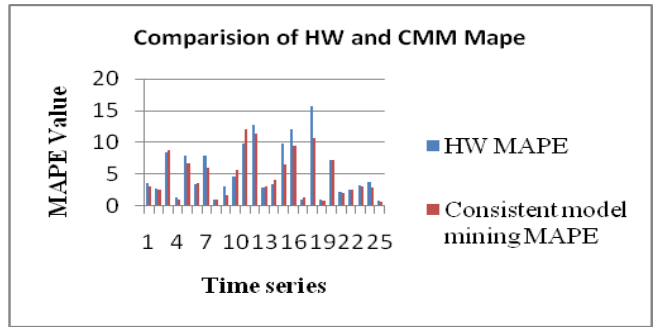

Figure 9. Comparison between Holt Winter MAPE and Consistent Model Mining MAPE

Table 1 presents the MAPE values and percentage of improvement as compared to the Holt-Winter forecasts for selected series. The numbers in the second column (labeled HW) are the MAPEs for the last two years using the Holt-Winter approach. Our approach is generally better than the simple approach.

The last row indicates the average percentage improvement in MAPE over HW forecasting averaged across all 25 series Using multiple decomposition methods. We achieved improvement in the forecasting accuracy and its performance. The result analysis shows that our approach is better than the HW, because from the above results we can say that at least 60% results are better than HW method.

Table 1. Comparison of CMM MAPE with HW MAPE

| Sr. No. | Series | HW MAPE | Consistent model mining MAPE | % of Improvement |
|---|---|---|---|---|
| 1 | Abraham | 3.53 | 2.97 | 15.82 |
| 2 | Beer | 2.64 | 2.43 | 7.87 |
| 3 | Dry | 8.3 | 8.76 | -5.64 |
| 4 | Equip | 1.23 | 0.95 | 22.517 |
| 5 | Fortif | 7.86 | 6.58 | 16.24 |
| 6 | Gasoline | 3.3 | 3.48 | -5.48 |
| 7 | Hsales | 7.87 | 5.93 | 24.57 |
| 8 | Merchandise | 0.87 | 0.88 | -1.92 |
| 9 | Motorparts | 2.87 | 1.56 | 45.63 |
| 10 | Newcar | 4.45 | 5.56 | -25.05 |
| 11 | Red | 9.79 | 11.99 | -22.53 |
| 12 | Rose | 12.66 | 11.38 | 10.11 |
| 13 | Shoe | 2.86 | 3.01 | -5.50 |
| 14 | Software | 3.25 | 4.06 | -25.06 |
| 15 | Spaper | 9.67 | 6.36 | 34.16 |
| 16 | Spark | 11.98 | 9.46 | 21.00 |
| 17 | Stores | 0.92 | 1.17 | -27.180 |
| 18 | Sweet | 15.6 | 10.54 | 32.43 |
| 19 | Total | 0.87 | 0.63 | 26.48 |
| 20 | Wine | 7.05 | 7.17 | -1.70 |
| 21 | Clothing | 2.1 | 1.85 | 11.65 |
| 22 | Furniture | 2.39 | 2.49 | -4.46 |
| 23 | Jewelry | 3.07 | 3.04 | 0.80 |
| 24 | sync2 | 3.64 | 2.78 | 23.59 |
| 25 | synd1 | 0.62 | 0.60 | 2.06 |
| | Average Improvement (All Series) | | | 6.81 |

## V  INTER AND INTRA TIME SERIES

A time series data set consists of sequences of values or events that change with time. Time series data is popular in many applications, such as the daily closing prices of a share in a stock market, the daily temperature value recorded at equal time intervals, and so on.

There are two way to represent time series data.

### A. Intra Time Series :

It can only reveal the co-relations of multiple time series at same time. It has little contribution to prediction, the ultimate destination of time series analysis. Taking stock data as an example, it can be found the rules like "If stock A goes up and stock B goes up then stock C will goes up on the same day (5%, 80%)"

There is no time difference between the items in the intra-transactional association rules, it cannot be found the rules like "If stock A goes up on the first day and stock B goes up on the second day then stock C will goes up on the third day (5%, 80%)". so it cannot be used to predict the trend of time series.

### B. Inter Time Series :

The inter-transactional association rules describe the relations between different transactions.

If we use it in multiple time series analysis, it can be analyzed the relations of different series at different time. So, it can be used to find the rules like "If stock A goes up on the first day and stock B goes up on the second day then stock C will goes up on the third day (5%, 80%)".

It is evident that in the example above the time dimension is added, the events in the rule occur in different time. They have sequential order in time dimension. If in an association rule "X(0) => Y(2)", event Y occurs two days after event X, then the rule will have the ability of prediction. According to this rule, while event X occurs some day, we can infer that event Y will most likely occur two days after.

The inter-transactional association rules can be viewed as an extension of traditional intra transactional association rules.

The mining of inter-transactional association rules is much complex than traditional association rules, it challenges the efficiency of mining algorithms.

In time series analysis, intra-time series can only reveal the correlations of multiple time series at same time. It is difficult to forecast the trend of time series, so is best to use for inter time series mining to find trends in different type of series. The mining problem of inter-transactional association rules in time series with which the trend can be forecast by the time difference between the prerequisite and the consequent in a rule.

#### a. Sales Frequency Difference (SFD) :

Her we define the steps required to calculate the SFD between two sales time series. Retail organization may have items whose units of sale differ very vastly in scale. For example, unit sales for Furniture sets may be in hundreds per month as compared to Clothing, that may be sold in thousands or groceries may be sold in lacks per month.

Instead of looking at the volumes of increase or decrease per month we use only the percentage of difference as a measure. We only use the quantity in percentage by which the successive values in the time series vary. Since percentage is independent of the scales of original series, they are comparable.

Let's take two series and computer the distance difference as P and Q, $P_i$ and $Q_i$ indicate the values of the series at point i.

1. Initialize SFD(P,Q) = 0

2. Start i=1, same starting for both series, $P_i$ and $Q_i$.
3. Next point, i+1 compute the movement from I, ($|P_i - P_{i+1}|$, $(Q_i - Q_{i+1}|$). The difference is recorded as up, down and flat, based on whether the sales figure went up, down or remained the same. For small changed the movement is flat.
4. We record the percentage of change from points 'i' to 'i+1' for both the series, P and Q.
5. If the movement for both the series from point 'i' to 'i+1' are not the same, we increment or decrement difference SFD (P,Q) by1.

6. If frequency is same for both the series, that is P, Q recorded the same movement; we check the percentage of change from the previous point. If the difference is within the permissible threshold, we consider that both the series have similar pattern, and if the change is beyond the permissible threshold we increment the count of SFD (P, Q) by 1.
7. Repeat steps 3 to 6 for the next points up to point 'n'
8. End

If the frequency pattern is same we can say these both series are having same pattern so the same experts may work for both series.

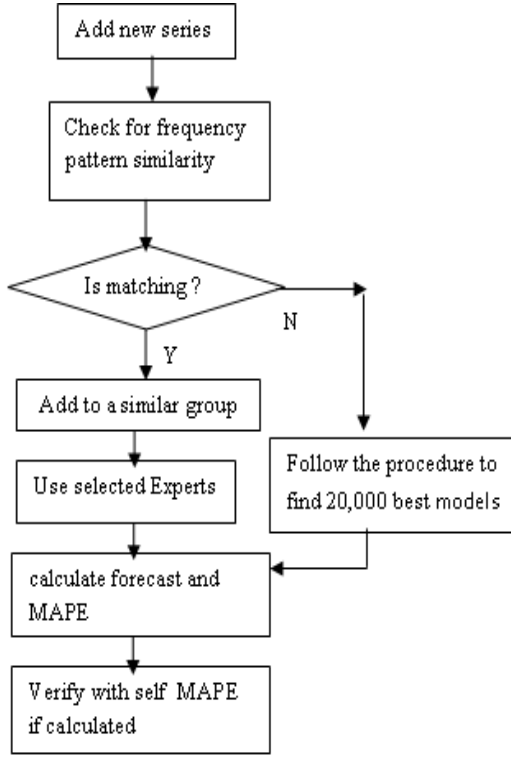

Figure 10. Flow of Inter time series mining

## V. EXPERIMENT RESULTS

**Experiment Results of Inter time series Mining:**
Using Furniture Best models we have calculated same frequency group MAPE in Table 2.

Table 2. Using Furniture Best Model group MAPE

| Sr. No. | Time series | Inter time Series MAPE |
|---|---|---|
| 1 | Software | 4.156712 |
| 2 | Abraham | 3.02555 |
| 3 | clothing | 2.22711 |
| 4 | Gasoline | 3.4954 |
| 5 | Newcar | 5.735 |

Using Synb1 Best models we have calculated same frequency group MAPE in Table 3.

## VI. SUMMARY AND CONCLUSION

Combining models reduces the risk in place of trusting a single forecast for sales forecasting. The forecasting accuracy is also improved by combining model. After decomposition a set of huge experts has achieved, these experts are used to find consistent experts using CHARM algorithm of association mining.

The forecast value is calculated by combining selected consistent model for a series in test phase(last 24 point of a series).The good and bad both experts has achieved but the results shown here are for good experts. The experiments have done on different values for support, confidence and also training set sizes.

The forecasted value is then compared with Holt Winter forecasted values and shown that our results are better than it.

Finally, we found the time series which are same in frequency, and some other factors are also there. The inter time series concept is working with the same frequency pattern time series. So the best model of one series of that group is best for other time series of the same group, as we have shown the result in Table 2 and Table 3.In future one can work with the bad experts.

Table 3. Using Synb1 Best Model group MAPE

| Sr. No. | Time series | Inter time Series MAPE |
|---|---|---|
| 1 | Synd1 | 0.6153 |
| 2 | Sync2 | 2.08081 |
| 3 | Sync1 | 0.68252 |
| 4 | Spaper | 6.79456 |
| 5 | Equip | 1.0161 |
| 6 | Total | 0.69356 |
| 7 | Motorparts | 1.55447 |
| 8 | Store | 1.19019 |

## VII. APPENDIX

The tables describe the various Trend, Seasonal and Irregular experts used in this research.

Table 1. Irregular Experts

| Expert ID | Expert Name | Expert ID | Expert Name | Expert ID | Expert Name |
|---|---|---|---|---|---|
| 1 | ARIMA (0,0,1)s | 13 | ARIMA (3,0,0)(1,0,0)s | 24 | Log ARIMA (1,0,1)s |
| 2 | ARIMA (0,1,0) | 14 | Linear Exponential | 25 | Log ARIMA (1,1,0) |
| 3 | ARIMA (0,1,1) | 15 | Linear Trend AR1 | 26 | Log ARIMA (1,1,2) |
| 4 | ARIMA (0,1,1)(1,0,0)s NOINT | 16 | Linear Trend AR2 | 27 | Log ARIMA (2,0,0) |
| 5 | ARIMA (0,1,1)s NOINT | 17 | Linear Trend AR3 | 28 | Log ARIMA (2,0,0)(1,0,0)s |
| 6 | ARIMA (1,0,0) | 18 | Log ARIMA (0,0,1)s | 29 | Log (3,1,1) NOINT |
| 7 | ARIMA (1,0,0)s | 19 | Log ARIMA (0,1,0) | 30 | Log Linear Exponential |
| 8 | ARIMA (1,0,1)s | 20 | Log ARIMA (0,1,1)(1,0,0)s NOINT | 31 | Log Linear Trend AR1 |
| 9 | ARIMA (1,1,0) | 21 | Log ARIMA (0,1,1)s NOINT | 32 | Log Linear Trend AR2 |
| 10 | ARIMA (1,1,2) | 22 | Log ARIMA (1,0,0) | 33 | Log Linear Trend AR3 |
| 11 | ARIMA (2,0,0) | 23 | Log ARIMA (1,0,0)s | 34 | Random |
| 12 | ARIMA (2,0,0)(1,0,0)s | | | | |

Table 2: Trend Experts

| Expert ID | Expert Name | Expert ID | Expert Name | Expert ID | Expert Name | Expert ID | Expert Name |
|---|---|---|---|---|---|---|---|
| 1 | ARIMA (0,1,0)(0,0,1)s | 11 | ARIMA (0,2,1) NOINT | 21 | ARIMA (1,1,1) NOINT | 31 | ARIMA (2,1,0) |
| 2 | ARIMA (0,1,0)(1,0,0)s | 12 | ARIMA (1,0,1) | 22 | ARIMA (1,1,2) | 32 | ARIMA (2,1,0)(1,0,0)s |
| 3 | ARIMA (0,1,0)(1,0,0)s NOINT | 13 | ARIMA (1,1,0) | 23 | ARIMA (1,1,2)(0,0,1)s | 33 | ARIMA (2,1,0)(1,0,0)s NOINT |
| 4 | ARIMA (0,1,0)(1,0,1)s | 14 | ARIMA (1,1,0)(0,0,1)s | 24 | ARIMA (1,1,2)(1,0,0)s | 34 | ARIMA (2,1,0) NOINT |
| 5 | ARIMA (0,1,1) | 15 | ARIMA (1,1,0)(1,0,0)s | 25 | ARIMA (1,1,2) NOINT | 35 | ARIMA (2,1,1) |
| 6 | ARIMA (0,1,1)(1,0,0)s NOINT | 16 | ARIMA (1,1,0)(1,0,0)s NOINT | 26 | ARIMA (1,2,0) | 36 | ARIMA (2,1,1) NOINT |
| 7 | ARIMA (0,1,1) NOINT | 17 | ARIMA (1,1,0)(1,0,1)s | 27 | ARIMA (1,2,0) NOINT | 37 | ARIMA (2,1,2) |
| 8 | ARIMA (0,1,2) | 18 | ARIMA (1,1,0) NOINT | 28 | ARIMA (1,2,1) | 38 | ARIMA (2,1,2) NOINT |
| 9 | ARIMA (0,1,2) NOINT | 19 | ARIMA (1,1,1) | 29 | ARIMA (1,2,1) NOINT | 39 | ARIMA (2,2,1) |
| 10 | ARIMA (0,2,1) | 20 | ARIMA (1,1,1)(0,0,1)s | 30 | ARIMA (2,0,1) | 40 | ARIMA (2,2,1) NOINT |
| Expert ID | Expert Name | Expert ID | Expert Name | Expert ID | Expert Name | Expert ID | Expert Name |
| 41 | ARIMA (3,1,0) | 53 | Log ARIMA (0,1,2) | 65 | Log ARIMA (1,1,1) NOINT | 77 | Log ARIMA (2,1,0) NOINT |
| 42 | ARIMA (3,1,0)(0,0,1)s | 54 | Log ARIMA (0,1,2) NOINT | 66 | Log ARIMA (1,1,2) | 78 | Log ARIMA (2,1,1) |
| 43 | ARIMA (3,1,0)(1,0,0)s | 55 | Log ARIMA (0,2,1) | 67 | Log ARIMA (1,1,2)(0,0,1)s | 79 | Log ARIMA (2,1,1) NOINT |
| 44 | ARIMA (3,1,0) NOINT | 56 | Log ARIMA (0,2,1) NOINT | 68 | Log ARIMA (1,1,2)(1,0,0)s | 80 | Log ARIMA (2,1,2) |
| 45 | Holt | 57 | Log ARIMA (1,1,0) | 69 | Log ARIMA (1,1,2) NOINT | 81 | Log ARIMA (2,1,2) NOINT |
| 46 | Log ARIMA (0,1,0)(0,0,1)s | 58 | Log ARIMA (1,1,0)(0,0,1)s | 70 | Log ARIMA (1,2,0) | 82 | Log ARIMA (2,2,1) |
| 47 | Log ARIMA (0,1,0)(1,0,0)s | 59 | Log ARIMA (1,1,0)(1,0,0)s | 71 | Log ARIMA (1,2,0) NOINT | 83 | Log ARIMA (2,2,1) NOINT |

Table 3: Seasonal Experts

| Expert ID | Expert Name | Expert ID | Expert Name | Expert ID | Expert Name |
|---|---|---|---|---|---|
| 1 | ARIMA(0,0,1)(0,1,1)s | 12 | ARIMA(2,1,0)(0,1,1)s | 23 | Log ARIMA(1,0,1)(0,1,1)s |
| 2 | ARIMA(0,0,2)(0,1,1)s | 13 | ARIMA(2,1,1)(0,1,1)s | 24 | Log ARIMA(1,1,0)(0,1,1)s |
| 3 | ARIMA(0,1,1)(0,1,1)s | 14 | ARIMA(2,1,2)(0,1,1)s | 25 | Log ARIMA(1,1,1)(0,1,1)s |
| 4 | ARIMA(0,1,1)s | 15 | ARIMA(3,0,0)(0,1,1)s | 26 | Log ARIMA(1,1,2)(0,1,1)s |
| 5 | ARIMA(0,1,2)(0,1,1)s | 16 | ARIMA(3,1,0)(0,1,1)s | 27 | Log ARIMA(2,0,0)(0,1,1)s |
| 6 | ARIMA(1,0,0)(0,1,1)s | 17 | Log ARIMA(0,0,1)(0,1,1)s | 28 | Log ARIMA(2,1,0)(0,1,1)s |
| 7 | ARIMA(1,0,1)(0,1,1)s | 18 | Log ARIMA(0,0,2)(0,1,1)s | 29 | Log ARIMA(2,1,1)(0,1,1)s |
| 8 | ARIMA(1,1,0)(0,1,1)s | 19 | Log ARIMA(0,1,1)(0,1,1)s | 30 | Log ARIMA(2,1,2)(0,1,1)s |
| 9 | ARIMA(1,1,1)(0,1,1)s | 20 | Log ARIMA(0,1,1)s | 31 | Log ARIMA(3,0,0)(0,1,1)s |
| 10 | ARIMA(1,1,2)(0,1,1)s | 21 | Log ARIMA(0,1,2)(0,1,1)s | 32 | Log ARIMA(3,1,0)(0,1,1)s |
| 11 | ARIMA(2,0,0)(0,1,1)s | 22 | Log ARIMA(1,0,0)(0,1,1)s | 33 | Holt Winter |